\documentclass[final,3p,times,authoryear,twocolumn]{elsarticle}

\usepackage{amssymb}
\usepackage{dsfont}
\usepackage{amsmath}
\usepackage{adjustbox}
\usepackage{hyperref}
\usepackage{rotating}
\usepackage{xcolor}

\newcommand{\bee}{\begin{equation}}
\newcommand{\eee}{\end{equation}}
\newcommand{\bea}{\begin{eqnarray}}
\newcommand{\eea}{\end{eqnarray}}

\journal{}

\begin{document}

\begin{frontmatter}



\title{Validation of methods to estimate the uncertainty of buildings energy savings in a controlled numerical setting and Bayesian energy signature with autocorrelated errors} 


\author{Léa Gondian and Thimothée Thiery} 

\affiliation{organization={Centre Scientifique et Technique du Bâtiment},country={France}}

\begin{abstract}
In the field of building energy efficiency, the measurement and verification (M\&V) of energy savings following energy efficiency measures often relies on the use of a calibrated statistical model. In order to  obtain reliable estimates, the estimation of uncertainties associated with this procedure is recognized as a crucial aspect of M\&V. Several approaches have been proposed in the literature to compute uncertainties but a recent work \citep{touzani2019evaluation} has raised serious doubts on their accuracy. This is especially the case for energy data with fine time resolution (hourly and daily), for which the issues associated with autocorrelations  become more important. This work focuses on this issue for daily data with three main contributions: (i) a detailed comprehensive review of several methods proposed in the literature for linear and non-linear regression models (including exact formulas for linear models, the ASHRAE 14 approximate formula and resampling approaches) ; (ii) the improvement of approaches based on Bayesian modeling for non-linear regression models with autocorrelated residuals ; (iii) a large scale numerical test of the different approaches based on a synthetic dataset containing thousands of stochastic thermal dynamic simulations. Our results permit to test the accuracy of different approaches in a controlled numerical setting. In contrast with \citep{touzani2019evaluation}, we find in particular that our Bayesian approach leads on our dataset to consistent estimates of uncertainties. This suggests that the proposed Bayesian approach offers a convincing method to estimate uncertainties in non-linear regression models with autocorrelated residuals often encountered M\&V. 
\end{abstract}




\end{frontmatter}




\section{Introduction}\label{sec1}

\subsection{Measurement and verification and scope of this work}

In the field of building energy efficiency, the accurate measurement and verification (M\&V) of energy savings following retrofit projects is crucial to ensure that energy efficiency measures (EEMs) meet their objectives. The energy consumption of real world buildings being affected by many varying factors (weather data, occupancy...), M\&V often relies on statistical approaches to cancel out in the savings estimates the effects of these varying conditions. As emphasized by the main references for M\&V such as the international performance measurement and verification protocol (IPMVP) \citep{ipmvp_core_concepts} and ASHRAE Guideline 14 \citep{ashrae14}, an accurate estimation of uncertainties associated with this statistical procedure is crucial for a correct interpretation of the results. This is in particular true when savings are estimated using meter data for the consumption of the whole building (IPMVP option C), as is often the most appropriate for building retrofits. The goal of this work is to review, extend and test existing approaches to compute uncertainties in IPMVP option C protocols. We will mainly focus on the case of data with daily time resolution and on the use of linear and non-linear regression techniques. While we focus on the field of building energy efficiency, the results should be of interest in other fields where M\&V can be applied, such as in industrial settings.

\begin{figure}[h]
    \centering
    \includegraphics[width=0.45\textwidth]{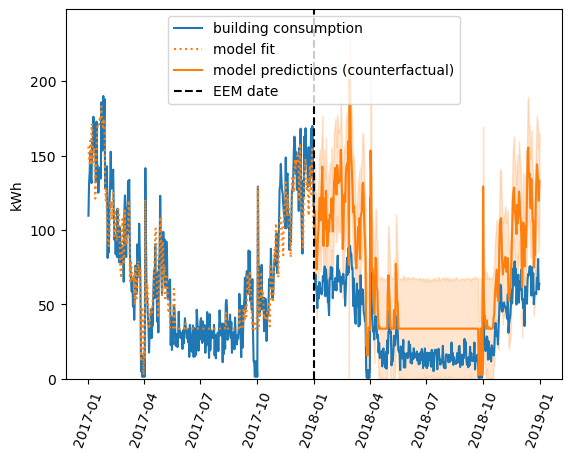}
    \caption{Example of M\&V workflow with synthetic daily energy consumption data. The building consumption pattern (blue line) changes after the implementation of the EEM (January 2018). A statistical model is calibrated to fit the building consumption before retrofit (baseline period, 2017) (orange dashed line). The model is used to predict what would have been the building consumption in the absence of retrofit during the reporting period (2018). The orange region represents the $95\%$ prediction interval  during the reporting period. The difference between the blue and orange curves during the reporting period is used to estimate the impact of the EEM. The modeling goal is to obtain an estimate of the effect and compute its uncertainty.}
    \label{fig:mandv}
\end{figure}

Diving now into more details, the goal is to compute the savings associated with some EEM implemented at a given time $t_{{\rm EEM}}$\footnote{In practice the EEM is not implemented during a single day but we can still distinguish a period before and after retrofit.}. In the framework of IPMVP option C the full daily energy consumption of the building was measured for (e.g.) a year before the intervention $E_d$, $d <t_{{\rm EEM}}$ (the baseline period), and also a year after the intervention $E_{d'}$, $d'  > t_{{\rm EEM}}$ (the reporting period). The raw savings are simply computed as $\sum_{d}E_d - \sum_{d'}E_{d'}$. As mentioned earlier, the issue with raw savings is that the building consumption is affected by external factors $X_d$ (e.g. the external temperature $\theta_d$, the global horizontal irradiation $I_d$...) which varies in between the baseline and reporting periods: $X_d \neq X_{d'}$. Therefore raw savings are affected by the variations in $X_d$ which cannot be attributed to the EEM. The idea is thus to use an adjustment mechanism to compute savings that takes into account these changes in external conditions. The solution to this issue is to calibrate a function $\hat{E}_d = f(X_d) $ representing building consumption as a function of external factors during the baseline period:
\bee \label{eq:1}
E_d = f(X_d) + \epsilon_d \, , 
\eee
where $\epsilon_d = E_d - \hat{E}_d$ are the residuals of the modeling. This calibrated function is then used to predict what would have been the consumption in the reporting period in the absence of EEM, $\hat{E}_{d'} = f(X_{d'}) $ (one often speaks of a "counterfactual" consumption). The adjusted savings are then computed as 
\bee
{\rm savings} = \sum_{d'} \left( \hat{E}_{d'} - E_d\right) = \sum_{d'} \left(  f(X_{d'}) - E_d\right) \, ,
\eee
see Fig.~\ref{fig:mandv} for illustration. This calculation being the result of the use of a statistical model, it is important to estimate its uncertainty. The goal of this paper is to review, improve and test uncertainty estimation approaches for M\&V modeling using daily data. We will focus mostly on linear and non-linear regression approaches but also briefly consider other techniques.

\subsection{Energy signature and change-points models}

Among the most studied models in M\&V are models representing the daily consumption $E_d$ of the building as a linear combination of the heating and cooling degree days at some given base temperatures $\theta_{h}$ and $\theta_{c}$. These are defined as $hdd_{d} = {\rm max} (0, \theta_h - \theta_d ) $ and $cdd_{d} = {\rm max}(0,  \theta_d  - \theta_c) $, with $\theta_d$ the average temperature during day $d$. The model can be written as

\bee
f(X_d) = \hat{E}_d = bl + \alpha_h \times hdd_d + \alpha_c \times  cdd_d \, . \label{eq:5points}
\eee
The three coefficients $bl$, $\alpha_h$ and $\alpha_c$ (respectively the base-load, heating slope and cooling slope) need to be calibrated to data and are usually assumed to be positive. The model is interpretable and the three terms are usually thought of as representing the heating consumption, the cooling consumption, and the other consumptions \citep{rabl1992energy,kim2022simplified}. This type of model is often referred to in the M\&V literature as a 5-points change-point model \citep{ashrae14}, see e.g.  \citep{fu2021review} for a recent review. From a formal point of view this is a linear model with $3$ parameters and $2$ features, $X_d = (hdd_d , cdd_d)$. Therefore, exact formulas existing for linear models can be used to obtain the coefficients, their uncertainty and any prediction intervals \citep{reddy2011applied}. The use of such formulas typically involve matrix calculations but an approximate simple formula was given in \citep{reddy2000uncertainty} and its use is recommended in ASHRAE Guideline 14 \citep{ashrae14}. While the situation seems simple in the case of linear models, let us note that already in that case the treatment of uncertainty is non-trivial when the residuals $\epsilon_d$ are autocorrelated \citep{ruch1999prediction,reddy2000uncertainty}. We will review this in details in Sec.~\ref{sec:overview}.

However, in practice  the base temperatures are not fixed and depend on the building. The M\&V practitioner usually optimizes $\theta_h$ and $\theta_c$ to obtain the best fit. Hence in that situation the model \eqref{eq:5points} has not $3$ free parameters but $5$ parameters and should be written as
\bee
f(X_d) = \hat{E}_d = bl + \alpha_h \times (\theta_h - \theta_d)^{(+)} + \alpha_c \times (\theta_d - \theta_c)^{(+)} \, , \label{eq:ES1}
\eee
where $(x)^{(+)}$ is a shorthand for ${\rm max}(0,x)$. Written in this way the model is often referred to as the "energy signature model" \citep{hammarsten1987critical}. Formally, it is a non-linear regression model with $5$ parameters and a single feature $X_d = \theta_d$. Hence, formulas for linear models are in theory not applicable here. The common practice then appears to first optimize $\theta_h$ and $\theta_c$ to obtain the best fit, to rewrite \eqref{eq:ES1} as \eqref{eq:5points} in order to be able to use the formulas for linear models, e.g. to compute the resulting uncertainty in energy savings. In essence this procedure {\it amounts to neglecting the uncertainty associated with the choice of the base temperatures}. As we will see in this paper a Bayesian approach, as already studied in \citep{lindelof2017Bayesian,lindelof2018Bayesian,shonder2012Bayesian} and pushed further here, permits to remove this approximation and to take into account all uncertainties.

We see from this discussion that the computation of uncertainties for one of the most simple and studied model in the M\&V literature relies on an uncontrolled assumption. Going one step further in the modeling, it is interesting to note that \eqref{eq:ES1}, given its interpretability as a heat balance equation for the thermal equilibrium of a building \citep{hammarsten1987critical,senave2020assessment}, can be extended to take into account more weather features as e.g. in \citep{rasmussen2020method}. In particular a version of \eqref{eq:ES1} that also takes into account the effect of solar gains on the heating load can be written as \citep{rasmussen2020method} 
\bee 
f(X_d) =  \hat{E}_d  = bl +
 \alpha_h \times (\theta_h - \theta_d - g\times I_{d})^{(+)} + \alpha_c \times (\theta_d - \theta_c)^{(+)}  \, . \label{eq:ES2} 
\eee
Here $I_{d}$ is a measure of the solar irradiation, typically taken as the average daily global horizontal irradiation at the building location, and $g$ an effective additional coefficient that needs to be calibrated. This model now has $6$ coefficients, $2$ features and appears "more non-linear" than \eqref{eq:ES1}. Exact matrix formulas available for linear models cannot be used but a standard industry practice in such situations is to use the approximate formula given \citep{ashrae14}. Since this approximate formula was derived from the study of linear models, clearly its use here relies on an uncontrolled assumption. One of the goals of this work is to develop and test an uncertainty quantification method that is appropriate for non-linear regression models with autocorrelated residuals. As noted in a recent review \citep{fu2021review} this indeed remains a largely open problem: "There are two 
main challenges in the uncertainty calculation for statistical building 
energy models: autocorrelation and non-linearity. There has not been a 
consensus for an analytical solution to these challenges.".  In this work we will show that a Bayesian approach permits to make progresses in that direction. We will mainly focus on \eqref{eq:ES2} but the approach can in principle be applied to arbitrary non-linear functions. Let us emphasize that, as we will show, the Bayesian approach studied in details here offers a clear and transparent way to estimate uncertainties for non linear regression models such as \eqref{eq:ES2}, but other approaches could be used as well. In particular in this paper we will briefly consider non-parametric approaches based on resampling algorithms such as conformal predictions \citep{angelopoulos2023conformal}, but other approaches exist. In particular we will not focus on frequentist approaches based on a linear approximation of the non-linear model around the calibrated parameters. The main motivation to focus on developing a Bayesian approach here is its versatility and transparency.

\subsection{Bayesian approaches in M\&V}

Given the importance of uncertainty quantification to obtain a reliable M\&V process, the application of the Bayesian paradigm to M\&V has been studied in the literature by several authors. It is even now mentioned as an option for uncertainty quantification in the latest ASHRAE 14 Guideline \citep{ashrae14}. This approach has been mainly tested in the framework of Gaussian Process Regression \citep{heo2012gaussian,maritz2018practical} as well as linear \citep{carstens2018Bayesian} and non-linear \citep{shonder2012Bayesian,lindelof2017Bayesian,lindelof2018Bayesian} regression models. However, investigation of the literature shows that the issue of autocorrelation of the residuals of the models has been largely overlooked except by \citep{shonder2012Bayesian}. This is intuitively problematic as it has been recognized as an important issue for M\&V in more traditional approaches \citep{reddy2000uncertainty} and there is no reason to believe that the Bayesian approach automatically fixes this issue: indeed we show here that it does not. Hence in this work we study in more details this issue and show that at present the approach of \citep{shonder2012Bayesian} is problematic and numerically unstable. We find a solution to this issue to obtain a Bayesian approach to uncertainty quantification for non-linear regressions with autocorrelated residuals that gives satisfying results.

\subsection{Model selection}

As already mentioned, an important question in M\&V is the issue of model selection: for example when to use \eqref{eq:ES2} instead of \eqref{eq:ES1}, or even a simpler model assuming only a base-load and a heating use $f(X_d) = bl +
 \alpha_h \times (\theta_h - \theta_d - g\times I_{d})^{(+)}$. In general the more features are used, the better the fit, but then appears the risk of "overfitting", when a model performs well on the training period (i.e. baseline) but provides a bad prediction in the test period (i.e. reporting). Standard IPMVP guidelines are rather vague on this issue. For example \citep{ashrae14} recommends: "In general, one would like a model selection procedure that is simple to apply and produces consistent,
repeatable results. Several procedures have been recommended to select the best regression results. In general,
these procedures calculate the results using several alternate models and then select the best model depending
on the value of R2 and CV(RMSE)". One advantage of a Bayesian workflow is that model selection can be performed using well grounded theoretical principles: in this work we will used the so-called Bayesian information criterion (BIC) (see e.g. \citep{rouchier2018solving}) and test it numerically.

\subsection{Testing M\&V methods in a controlled setting}

Finally, an important methodological point addressed in this work is the question of how to assess the accuracy of M\&V approaches. Investigation of the literature shows that there have been very few test of M\&V methods at large scale to assess their accuracy and, in particular, the validity of uncertainty predictions. An interesting initiative aiming at the large scale test of building energy forecasting methodologies was the great energy predictor III competition \citep{miller2020ashrae}. The goal there was to compare the performance of various building energy prediction methodologies using real data included in the open source dataset of the building data genome project \citep{miller2020building}. However, the focus of this competition was on the prediction of hourly energy consumption data without studying in detail the performance of the aggregated prediction on longer periods (e.g. one year) or the  uncertainty of the estimates, as is relevant in a standard M\&V setting. Another series of work has investigated the accuracy of the predictions \citep{granderson2014development,granderson2015automated,granderson2016accuracy} and of the estimation of the uncertainty \citep{touzani2019evaluation} for hourly and daily models using data from 537 real commercial buildings. In these works the models are calibrated on past data (training period) and the goal for the models is to predict the value of the consumption of some new data (testing period). The assumption is that the prediction and the real values should be as close as possible as the selected building did not undergo retrofit works in between the two periods: the savings should effectively be zero. These works were the basis for the development of an online tool \citep{evotesting} which allows developers of hourly M\&V models to test and compare the performance of their models to others. While extremely interesting, this approach has a limitation: it is difficult to ensure that building behavior remained unchanged between testing and training periods. For this reason the results can be hard to interpret. As noted in \citep{granderson2016accuracy} the results of this approach showed that models typically appeared to have a small positive bias. However, the authors could not decide whether this was due to a modeling issue or to a data issue: "for the majority of cases there was a tendency of a bias toward overpredicting
the energy use (NMBE negative). However, this may be a result of actual decreases in building energy use over time,
as opposed to a characteristic of the models. Further research is
needed to explore this premise." In the same way, one striking conclusion of \citep{touzani2019evaluation} is that the current estimation methodologies for uncertainties seem to be unreliable: the best tested methodology provided an estimate of a 95\% confidence interval for the yearly consumption that contained the true value only 71\% of the time. This result was attributed in \citep{touzani2019evaluation}  to the way autocorrelation is taken into account which was deemed as unreliable, especially for hourly models but also for daily models. However, this conclusion could again actually be affected by an issue with the data.

 To avoid this issue, in this work we propose an alternative procedure to test M\&V approaches using a large number of stochastic thermal dynamic simulations of randomly selected French residential buildings. This approach is not necessarily better than an approach using real data but complements it. Indeed, on one hand one cannot expect a simulation to reproduce all the complexity of real world data. On the other hand, simulated consumption data offer a testbed for which methods can be tested in an unambiguous manner. We therefore expect that a reliable M\&V method  should at the minimum give satisfying results on simulated data: our numerical test can thus be seen as a consistency test that permits to point out approaches which need to be refined. As we will show we indeed find methods that perform well on our synthetic data, and other methods which fail at providing fully consistent uncertainty estimates. Clearly, methods in the second category cannot be expected to give satisfying results on real world data.

 \subsection{Summary of questions addressed in this work and outline}\label{subsec:summary_of_questions}

Let us finish this introduction with a summary of some of the questions addressed in this work 
\begin{enumerate}
    \item Do the existing approaches that exist for linear models such as \eqref{eq:5points} appear valid when tested in a fully controlled numerical setting ? 
    \item How to take into account the uncertainty in the base temperatures in \eqref{eq:ES1} ? What is the effect of neglecting this uncertainty (the standard industry practice) ?
    \item More generally, how to compute uncertainty appropriately for non-linear regression models such as \eqref{eq:ES2} ?
    \item What is the accuracy of the simplified formula of \citep{ashrae14} when applied to linear and non-linear models ?
\end{enumerate}

The outline of this work is as follows. In Sec.~\ref{sec:overview} we review existing approaches to estimate the uncertainties in the M\&V context. In Sec.~\ref{sec:Bayes} we discuss in details our Bayesian workflow and how to deal with autocorrelated residuals in a Bayesian setting. In Sec.~\ref{sec:jdds} we discuss our methodology to build a synthetic dataset suited to the testing of M\&V methodologies. In Sec.~\ref{sec:numerical_tests} we perform a numerical test of different methodologies using our synthetic dataset. This will permit to compare the accuracy of the different approaches in a simple and controlled numerical setting. We will conclude in Sec.~\ref{sec:conclusion} with a summary of our findings and highlight open questions left for future works.

\section{Overview of existing (non Bayesian) approaches for uncertainty quantification}\label{sec:overview}

In this section we review some existing approaches outside of Bayesian ones, first focusing on linear models (Sec.~\ref{subsec:linear_models}), then outlying briefly modern approaches based on cross validation in Sec.~\ref{subsec:cv}

\subsection{Uncertainty computations in linear models}\label{subsec:linear_models}

Linear models are such that $f(X_t)$ in \eqref{eq:1} is assumed to be a linear function $f(X_t) = \sum_{i =1}^{n_f} \beta_i X_t^{i}=   X_t \cdot\beta  $, where $\beta = (\beta_1 , \cdots , \beta_{n_f} )'$ is the (column) vector of parameters that should be estimated, and $X_t = (X_t^{1} , \cdots , X_t^{n_f})$ is the (line) vector of features at times $t$ (which include the intercept term). In matrix form the full model can be written as

\bea
E = X \cdot \beta + \epsilon
\eea
where $E$ is the vector of consumptions, $X$ is a matrix of size $(n_{train} , n_f)$
and $\epsilon$ is the noise. Once the value for the parameter $\beta$ is fitted as $\hat{\beta}$, 
the prediction during the observation period is $\hat{E}_{obs} = X_{obs} \cdot \hat{\beta}$ and the issue is to compute a confidence interval for the total consumption during the observation period $\hat{E}_{tot} = \sum_{t = 1}^{n_{obs}} \hat{E}_{obs, t}  = \mathds{1} \cdot \hat{E}_{obs} $ where $\mathds{1} $ is a line vector with all inputs equal to $1$. Typically this implies estimating the variance-covariance matrix of prediction errors $\Sigma_{t,t'} = <(E_{obs,t} -\hat{E}_{obs,t})(E_{obs,t'} -\hat{E}_{obs,t'})>$ (a $n_{obs} \times n_{obs}$ matrix), which mainly involves computing the variance-covariance matrix of fitted coefficients $\Omega_{i,j} = <(\hat{\beta}_i - \beta_i)(\hat{\beta}_j - \beta_j)>$ (a $n_{f} \times n_{f}$ matrix). Above and everywhere, as usual in the frequentist approach, $<>$ denotes the average over the realization of the noise term $\epsilon$.

\subsubsection{Ordinary least square with uncorrelated residuals}\label{subsubsec:ols}

In the simplest approach, the residuals are assumed to be Gaussian, independent and identically distributed (iid) with standard deviation $\sigma$. The parameters $\beta$ and $\sigma$ are estimated using an ordinary least square approach (OLS) as 
\bee \label{eq:ols_beta}
\hat{\beta} = \hat{\beta}_{OLS} =  (X'X)^{-1} X' E
 \eee
 and 
 \bee \label{eq:ols_sigma}
 \hat{\sigma}^2 = \frac{1}{n_{train} - n_f} \sum_i \hat{\epsilon_i}^2
 \eee
 with $\hat{\epsilon_i}= E_i -\hat{E}_i$. The variance-covariance matrix of the coefficients $\hat{\beta}$ is estimated as 
 \bee \label{eq:ols_var_beta}
 \hat{\Omega} = \hat{\sigma}^2 (X'X)^{-1} .
 \eee
 Finally the variance-covariance matrix $\Sigma$ of the $n_{train}$ predictions errors $\hat{\epsilon}_{obs,i}= E_{obs,i} -\hat{E}_{obs,i}$  is  estimated as

\bea \label{eq:cov_mat_idd}
\hat{\Sigma} && =  \left(  X_{obs}   \hat{\Omega} X_{obs}'  +  \hat{\sigma}^2 \mathds{I} \right) \nonumber \\
&& = \hat{\sigma}^2 \times \left(  X_{obs}  (X'X)^{-1}   X_{obs}'  +  \mathds{I} \right)\, 
\eea 
with $\mathds{I}$ the identity matrix of size $(n_{obs},n_{obs})$. The variance of the total consumption in the observation period is then given by the sum of all the elements of this matrix: $Var(\hat{E}_{tot}) = \sum_{i,j} \Sigma_{ij}$. A confidence interval on $E_{tot}$ is then obtained using that $\hat{E}_{tot} - E_{tot}$ should be distributed as a centered Student distribution with $n_{train}-n_f$ degrees of freedom and variance given by the above.

It is worth emphasizing that even in this simple modeling where the residuals are assumed to be uncorrelated, the prediction errors are in fact correlated since $\hat{\Sigma}$ has non-zero diagonal elements. The correlation in the prediction errors come from the first term in \eqref{eq:cov_mat_idd} which accounts for the uncertainty in the estimation of $\hat{\beta}$ (included in $\hat{\Omega}$). Prediction errors are autocorrelated since all predictions share the same value of $\hat{\beta}$. On the other hand the fact that the second term in \eqref{eq:cov_mat_idd} is proportional to the identity and thus has zero non-diagonal elements is a consequence of assuming that the noise is not correlated.

\subsubsection{Modeling autocorrelations using an AR(1) process}\label{subsubsec:ruch}

A common issue appearing in regression approaches to time-series data and in particular in M\&V analysis is when the residuals of the modeling appear to be correlated in time. Uncorrelated residuals should satisfy $<\hat{\epsilon}_i \hat{\epsilon}_{i+l}> = 0$ for all values of the lag $l > 0$. On the contrary, autocorrelated residuals display non-zero correlations for lag $ l > 0$. Testing of autocorrelations in the residuals can be done using the Durbin-Watson test or more simply by plotting the autocorrelation function of the residuals\footnote{Assuming here in \eqref{eq:acf} that the residuals have zero mean.}:
\bea \label{eq:acf}
acf(l) = \frac{n_{train} \sum_{i = 1}^{n_{train}- l} \hat{\epsilon}_i \hat{\epsilon}_{i+l}}{(n_{train} - l ) \sum_{i = 1}^{n_{train}}\hat{\epsilon}_i^2} \, 
\eea
and in particular its value at lag $1$, often noted $\rho$
\bea
\rho = \hat{\rho}_{OLS} =acf(1) \, .
\eea

A common approach that has been studied in the M\&V context \citep{ruch1999prediction} is to assume that the residuals can be modeled as a stationary $AR(1)$ process, i.e. that $\epsilon_{t} = \rho \epsilon_{t-1} + \delta_t$ with $-1<\rho<1$ and $\delta_t$ an auxiliary Gaussian noise which is iid, centered around $0$ with variance $\sigma'$. In that case $acf(l) = \rho^l$ and $\epsilon_t$ is a stationary Gaussian process centered around $0$, with a covariance $<\epsilon_{t} \epsilon_{t+l} > = \frac{(\sigma')^2}{1-\rho^2} \rho^l$ that is geometrically decreasing when the lag $l$ increases. There is an important literature on the modeling of linear regression + autocorrelated noise. An issue in that context is that in general $\rho$ is unknown and several procedures have been studied to estimate it simultaneously with the other parameters.

Common approaches are the Cochrane-Orcutt and Prais-Winsten iterative procedures (see e.g. \citep{jaggia2008practical}) where first $\beta$ is estimated as before using OLS, $\hat{\beta} = \hat{\beta}_0 = \hat{\beta}_{OLS}$, assuming first no autocorrelations in the noise ($\rho = 0$). Then a first value of $\rho$ is estimated by computing the residuals of the fitted model, $\rho = \rho_0  = \rho_{OLS}$. The model is then fitted once more using generalized least square (GLS) assuming that $\rho$ is given by the previous estimate $ \rho_0$, leading to a new value of $\hat{\beta} = \hat{\beta}_1$. A new value of $\rho = \rho_1$ can be estimated by studying the residuals of the new modeling and the procedure is then iterated until the value of $\rho$ and $\hat{\beta}$ converges. The statistical literature on this kind of modeling suggests that in the end the estimate of the coefficients $\beta$ obtained by OLS is correct (in the sense that is not biased), but the estimate of its variance is not correct, which in turns has an impact on  prediction uncertainties.

However, it was noticed in the M\&V context \citep{ruch1999prediction} that this procedure was leading to deceptive results, with the new estimate of $\beta$ obtained this way  corresponding to modeling having lesser predictive power compared with the OLS approach. The suggestion of \citep{ruch1999prediction} was therefore to keep the original OLS estimate of $\hat{\beta}$ and only to modify the uncertainty calculations to take into account the AR(1) hypothesis for the noise. This leads to a hybrid model where 

\bea
\hat{\beta} = \hat{\beta}_{OlS} , 
\eea
\bea
\hat{\sigma}^2 = \frac{\sum_i \hat{\epsilon}_i^2}{\sum_i (M \Psi)_{ii}} \, 
\eea
where $\Psi_{ij}$ is the $(n_{train},n_{train})$ autocorrelation matrix of the residuals that is assumed to take the AR(1) form $\Psi_{ij} = \rho^{|i-j|} $ and $M = \mathds{I} - X(X'X)^{-1}X'$. The variance-covariance matrix of $\hat{\beta}$ and of the predictions are then estimated\footnote{The expression of $\Sigma$ here neglects possible autocorrelations between the training and observation period. The complete expressions are given in \citep{ruch1999prediction}. For clarity of the exposure and comparison with other techniques we have rewritten here the expressions in \citep{ruch1999prediction} in a slightly different form, also correcting a missprint present in \citep{ruch1999prediction}. $P^{-1}$ in Eq.(11)-(12) there should indeed be replaced by $P'$.} as
\bea
\hat{\Omega} = \hat{\sigma}^2 (X'X)^{-1}X' \Psi X (XX')^{-1} \ ,
\eea
\bea \label{eq:cov_mat_ruch}
\hat{\Sigma} = \left(X_{obs}  \hat{\Omega} X_{obs}' + \hat{\sigma}^2  \Psi_{obs} \right) \ ,
\eea 
with $\Psi_{obs}$ the $(n_{obs},n_{obs})$ matrix that takes again the AR(1) autocorrelation form $\Psi_{obs,ij} = \rho^{|i-j|}$. The rest of the procedure to compute the uncertainty of a sum of predictions is then identical, using \eqref{eq:cov_mat_ruch} instead of \eqref{eq:cov_mat_idd}. Notice that compared to the OLS case, correlations between the prediction errors come from two sources: both terms appearing in \eqref{eq:cov_mat_ruch} have non-zero diagonal elements. As in the OLS case, the first term accounts for the uncertainty on $\hat{\beta}$. The second term, which represents the residuals autocorrelation function, is also in that case non-diagonal.

\subsubsection{An approximate formula (ASHRAE 14)}\label{subsubsec:ashrae14}

In order to avoid the above matrix calculations, an explicit formula approximating \eqref{eq:cov_mat_idd} and \eqref{eq:cov_mat_ruch} was given in \citep{reddy2000uncertainty}. The formula amounts to estimate the variance of the total predicted consumption in the observation period as 
\bea \label{eq:ashrae14}
  \Delta \hat{E}_{tot} &&  = \sqrt{Var(\hat{E}_{tot})}  \\ 
&& = 1.26 \frac{\overline{\hat{E}_{obs}}}{\overline{E_{train}}}\hat{\sigma} \sqrt{\frac{n_{train}}{n_{train}'} \left(1 + \frac{2}{n_{train}'} \right) n_{test}}
\nonumber
\eea
where $\overline{\hat{E}_{obs}}$ is the average predicted daily consumption in the observation period, $\overline{E_{train}}$ is the average measured daily consumption in the training period and $n_{train}'$ is the {\it effective number of observations in the training period}, given by
\bea
n_{train}'  = n_{train} \times \frac{1-\rho}{1+\rho} \, .
\eea
This number can be interpreted as the number of effectively independent observations in the training period, which, due to autocorrelations, is smaller than $n_{train}$ (for the usual case of positive autocorrelations $\rho >0$). Using  by definition $\hat{E}_{tot} =  n_{train}\overline{\hat{E}_{obs}} $, we obtain that the relative uncertainty is given by (assuming $n_{train}' >>2$ for simplicity)
\bea\label{eq:ashrae14-2}
\frac{\Delta \hat{E}_{tot}}{\hat{E}_{tot}} \simeq 1.26 \frac{\hat{\sigma}}{\overline{E_{train}}} \times \frac{1}{\sqrt{n_{test}}} \times \sqrt{\frac{1+\rho}{1-\rho}} \, .
\eea
This formula is interesting because it permits to simply highlight strategies to reduce uncertainties in M\&V: reducing $\hat{\sigma}$, increasing $n_{test}$ and having $\rho$ as small as possible. Reducing $\hat{\sigma}$ typically means having a model that provides a better fit, which can be done e.g. by adding features in the modeling. Increasing $n_{test}$ can be done by using a longer observation period or by increasing the frequency (e.g. going from using monthly to daily data or hourly data). However, note that increasing the frequency will also have some detrimental effects: it typically increases $\hat{\sigma}$ and $\rho$ as autocorrelation become more and more significant. With a too high frequency, it is also to be expected that the modeling becomes not representative of the reality of the building and that the hypothesis of AR(1) autocorrelations of the noise will break down. This highlights that given the modeling tools at our disposal, there probably is an optimum frequency to be considered when performing M\&V. In particular the advantages of using hourly modelings instead of daily modelings remain unclear as of today \citep{touzani2019evaluation}. In this work we focus on daily modelings and leave this interesting question for future works.

\subsubsection{Newey-West approach}\label{subsubsec:newey-west}

An important assumption underlying formulas in Sec.~\ref{subsubsec:ruch} and Sec.~\ref{subsubsec:ashrae14} is the hypothesis that the noise autocorrelations are given by that of an AR(1) process. To alleviate this assumption, Newey and West proposed in \citep{newey1986simple} an estimator of the variance-covariance matrix of $\hat{\beta}$ that is in principle robust to heteroskedasticity and autocorrelation effects (HAC estimator). Following this approach $\hat{\beta} = \hat{\beta}_{OlS} $ and the variance covariance matrix of coefficients is estimated as
\bea
\hat{\Omega} = (X'X)^{-1} M_{NW} (X'X)^{-1} \, ,
\eea
with
\bea
\hspace{-1cm} M_{NW} =&& \frac{n_{train}}{n_{train}-n_f} \left[ \sum_{i=1}^{n_{train}} \hat{\epsilon}_i^2 (X'_iX_i)  \right. \\
+ &&  \left. \sum_{l=1}^L \sum_{i=1}^{n_{train}-L} \left(1 - \frac{l}{L+1} \right) \hat{\epsilon}_i \hat{\epsilon}_{i+l} \left(   X'_iX_{i+l} + X'_{i+l} X_i \right)\right] \nonumber \, ,
\eea
Where $L \geq 1$ in the above expression is an integer that corresponds to the number of lags taken into account to model autocorrelations. In this paper we use the standard choice $L \sim n_{train}^{1/4}$. The covariance of the residuals for the predictions is in the same way estimated as $\Psi_{NW,i j} = \varphi_{NW}(|i-j|)$ where 
\bea
\varphi_{NW}(l) = \frac{1}{n_{train}-n_f}  \sum_{i=1}^{n_{train}-L} \left(1 - \frac{l}{L+1} \right) \hat{\epsilon}_i \hat{\epsilon}_{i+l}
\eea
And both are combined to obtain the variance-covariance matrix of the predictions 
\bea \label{eq:cov_mat_nw}
\hat{\Sigma} = \left(X_{obs}  \hat{\Omega} X_{obs}' +\Psi_{NW}  \right) \ .
\eea 
Although the option of using Newey-West standard errors has been cited in \citep{ipmvp_uncertainties}, we are not aware of this procedure having been tested in the scientific literature on M\&V.

\subsection{Non-parametric approaches using resampling}\label{subsec:cv}

The previous section has focused on reviewing analytical formulas that can be obtained in cases where the regression model $f(X_t)$ is a linear function. Another set of approaches that do not rely on a specific shape for $f(X_t)$ are {\it resampling} approaches. In such approaches the predictive accuracy of a modeling is estimated by training the modeling on a new training set consisting of the true training set minus a "hidden" part, and then evaluating the predictive accuracy of the modeling on the hidden part. This is at the core of conformal predictions approaches (see e.g. \citep{angelopoulos2023conformal}) which have received a lot of attention recently in the machine-learning community, but we note that applying it to time-series data poses particular challenges which are the object of current research \citep{angelopoulos2023conformal}. Conformal predictions do not have seem to be tested thoroughly in the M\&V context. In \citep{touzani2019evaluation} a closely related approach was proposed where a standard error on predictions $\hat{\sigma}_{CV}$ is estimated using $5-fold$ cross-validation (CV) where the training set is split in $5$ equal parts, and the model is then fitted $5$ times on all part but ones and the residuals are computed on the hidden part each time. In \citep{touzani2019evaluation} the authors discusses of some difficulties of performing this approach with autocorrelated time-series data and we refer to their work for more details. In the end they estimate the variance of $\hat{E}_{tot}$ as
\bea \label{eq:cv}
Var(\hat{E}_{tot}) \simeq n_{train} \times  \hat{\sigma}_{CV}^2 \times \frac{1+\rho}{1-\rho} \, 
\eea
where the factor of $\frac{1+\rho}{1-\rho}$ is added to correct for the effects of autocorrelations in the same way as in \eqref{eq:ashrae14-2}. This formula was only tested in \citep{touzani2019evaluation} with results similar to \eqref{eq:ashrae14-2} and overall deceptive results. As mentioned in the introduction, this could be due to biases introduced by the dataset used in \citep{touzani2019evaluation}. Overall the use of resampling approaches in M\&V does not appear to have been thoroughly investigated and this could deserve future work. In this paper we will simply investigate the accuracy of \eqref{eq:cv} without attempting to improve it. 

\section{Bayesian energy signature model with autocorrelated residuals} \label{sec:Bayes}

We now explain our Bayesian approach to compute uncertainties for non-linear regression modelings of the form \eqref{eq:ES2}. The reader is assumed to have some familiarities with Bayesian approaches and can refer to \citep{mcelreath2018statistical} for generalities and to \citep{lindelof2017Bayesian,shonder2012Bayesian} for a closely related application to energy signature models.

\subsection{A Bayesian energy-signature modeling with AR(1) correlated noise} 

The full model can be written as 
\bea
&& \hspace{-1.5cm}  f(\kappa \, ; X_d)  = bl +
 \alpha_h \times (\theta_h - \theta_d - g\times I_{d})^{(+)} + \alpha_c \times (\theta_d - \theta_c)^{(+)}  \, \nonumber \\
 && \hspace{-1cm}  E_d = f(X_d) + \epsilon_d \, ,
 \label{eq:ES_w_sun} 
\eea
with, in order to model the effect of autocorrelations, $\epsilon_d$ assumed to take the form of a stationary AR(1) process with parameters $\rho$ and $\sigma'$, i.e.
\bea  \label{eq:EQ_AR1_def}
&& \epsilon_1 \sim {\cal N}(0, \sigma) \nonumber \\  \, 
&& \epsilon_t = \rho \times \epsilon_{t-1} + \eta_t \text{ for $t \geq 2$}  \nonumber  \\ \, 
&& \eta_t \sim {\cal N}(0, \sigma') \, 
\eea
with the $\eta_t$ all iid and $\sigma^2 = \frac{(\sigma')^2}{1-\rho^2}$ to ensure that the noise is stationary (i.e. $\forall t, \epsilon_t \sim {\cal N}(0, \sigma)$). Here and everywhere $\sim$ means distributed as and ${\cal N}(\mu , \sigma)$ represents a standard normal random variable with parameters $\mu$ and $\sigma$. The notation $\kappa = (bl , \alpha_h, \theta_h, g, \alpha_c, \rho , \sigma')$ was also used to consider all model parameters simultaneously. In the Bayesian setting the model parameters $\kappa$ are also assumed to be random and distributed with a {\it prior} distribution $P_{\rm prior}(\kappa)$. The posterior probability distribution of $\kappa$ given the data ${\cal D} = (X_t, E_t)$ can then be computed using Bayes formula:
\bea
P( \kappa | {\cal D})  = \frac{ P( {\cal D} | \kappa) P_{\rm prior}(\kappa)}{N} \, ,
\eea
where $P( {\cal D} | \kappa) $ is the probability of observing the data points ${\cal D}$ given the parameters $\kappa$ and $N$ is a normalization constant. Using the logarithm, the above equation is rewritten as 
\bea
{\cal L}_{\rm post}(\kappa | {\cal D}) = {\cal L}_{\rm like}( {\cal D} | \kappa ) + {\cal L}_{\rm prior}(\kappa ) + N' \, ,
\eea
where we have introduced the log-posterior, log-likelihood and log-prior respectively given by ${\cal L}_{\rm post}(\kappa | {\cal D}) = \log(P( \kappa | {\cal D}))$, ${\cal L}_{\rm like}( {\cal D} | \kappa ) = \log(P( {\cal D} | \kappa))$, ${\cal L}_{\rm prior}(\kappa )  = \log( P_{\rm prior}(\kappa))$. $N' = \log(N)$ is a normalization constant and can be ignored in the following.

\subsection{Links with the existing literature}

The above modeling belongs to the family of Bayesian regression models with autocorrelated errors. We note that such models have been much studied in the statistics and econometrics literature, see e.g. \citep{greenberg2012introduction}. However, to our knowledge, their use in a M\&V context has only been studied in \citep{shonder2012Bayesian} and very recently in \citep{rouchier5333168bayesian}. In fact the above modeling generalizes that of \citep{shonder2012Bayesian} by introducing solar radiation in the model as done in \citep{rasmussen2020method}. As we show below in Sec.~\ref{subsec:instab}, the use of models with AR(1) autocorrelated errors in a M\&V context requires special care to avoid an instability in the calibration of parameters. Solving this instability is the main theoretical novelty of our work compared with \citep{shonder2012Bayesian}. In \citep{rouchier5333168bayesian} the author also study the impact of the autocorrelation of residuals for M\&V with Bayesian energy signature models. The modeling approach used in this second work mainly differs from ours by two aspects : (i) while in our work autocorrelations are modeled using and AR(1) process, in \citep{rouchier5333168bayesian} the author study different autocorrelation structures using moving-average models ; (ii) while in our work we mostly study the Bayesian approach using the Laplace approximation, \citep{rouchier5333168bayesian} uses a MCMC sampling of the posterior probability. Finally, our work also differs from \citep{shonder2012Bayesian} \citep{rouchier5333168bayesian} by the fact that we perform extensive statistical tests of our approach using a controlled numerical setting. Additionally let us note that our modeling is also closely related to the one studied recently in \citep{smertinas2025estimation} in the context of the evaluation of intrinsic building performance parameters (e.g. the overall heat loss coefficient). Indeed in \citep{smertinas2025estimation} the authors  study the effect of adding an autoregressive term (different from the one considered here) on the values of the fitted coefficients of the modeling. It would be interesting to assess if the improvements made in our work could also have implications in the context of \citep{smertinas2025estimation}. 

\subsection{Exact expression of the log-likelihood}

An exact expression for the log-likelihood can be obtained using the above expressions (see also e.g. \citep{beach1978maximum}). Using $\epsilon_t = E_t - f(X_t)$ one obtains
\bea
&& {\cal L}_{\rm like}( {\cal D} | \kappa )  = - n_{train} \log(\sigma') + 0.5 \log(1-\rho^2) \nonumber \\
&&  (1-\rho^2) \frac{\epsilon_1^2}{2 (\sigma')^2} -0.5 \frac{\sum_{t=2}^{n_{train}}(\epsilon_{t} - \rho \epsilon_{t-1})^2}{(\sigma')^2}
\eea

\subsection{Assumed priors}

The priors used in this work assume the form
\bea
{\cal L}_{\rm prior}(\kappa ) = {\cal L}_{\rm prior}(\sigma) + {\cal L}_{\rm prior}(\theta_h , \theta_c) \, 
\eea
with 
\begin{itemize}
    \item 
    ${\cal L}_{\rm prior}(\sigma) $ given as in \citep{lindelof2017Bayesian} as an inverse prior (here $\sigma^2 =\frac{(\sigma')^2}{(1-\rho^2)}$)
\bea
{\cal L}_{\rm prior}(\sigma) = - \log(\sigma)
\eea
\item 
${\cal L}_{\rm prior}(\theta_h , \theta_c)$ taken to enforce $\theta_h \in [5,25] $, $\theta_c \in [10,35]$ and $\theta_h < \theta_c$ as
\bea
{\cal L}_{\rm prior}(\theta_h , \theta_c) & =&  2 \log(\theta_h -5) + 2 \log(25-\theta_h)  \nonumber \\
&+& 2 \log(\theta_c -10) + 2 \log(35-\theta_c) \nonumber \\
&-& 100 \times (\theta_c - \theta_h)^{(+)}
\eea
This prior structure is mainly used to constrain the base temperatures to physical values, and avoid instabilities with cases where cooling and heating are simultaneously occurring ($\theta_c <\theta_h$), which is a known instability that can make the energy signature model non-invertible \citep{rabl1992energy}.

\end{itemize}

\subsection{Calibration strategy}

The previous section completely defines the non-linear Bayesian modeling that we study in this work. In the Bayesian approach the goal is then to obtain, given the data, the distribution of posterior distribution of $\kappa$, $P( \kappa | {\cal D})$. There are mainly two approaches to do so: use a Monte-Carlo Markov chain (MCMC) algorithm (or some generalization of it) to sample the posterior distribution directly (see e.g. \citep{mcelreath2018statistical}), or as in \citep{lindelof2017Bayesian}, use the Laplace approximation (see e.g. \citep{mackay2003information}) to approximate the posterior as a Gaussian distribution around its maximum. In this work we use this second approach that we detail below. We use this for practical simplicity and scalability of the approach as the Laplace approximation is faster and more easily automatized than a MCMC algorithm.  We have compared on a few examples the outcome of the two approaches and shown that they both lead to comparable results. However, the Laplace approach remains an approximation and the similarity between the results does not hold in full generality : MCMC generally represents a more rigorous approach to the problem and could be used instead.

In a first step the Laplace approximation approach involves computing the maximum a posteriori parameters defined as
\bea
\kappa^\star = {\rm argmax}_\kappa {\cal L}_{\rm post}(\kappa | {\cal D}) \, ,
\eea
The second step uses that the posterior distribution can be approximated as a multidimensional Gaussian distribution around $\kappa^\star$: 
\bea \label{eq:laplace1}
P( \kappa | {\cal D}) = {\cal N}(\kappa^\star , H_{\kappa^\star} ^{-1}) \, 
\eea
where the covariance matrix of the parameters is estimated as the inverse of the Hessian at the maximum:
\bea \label{eq:hessian}
H_{\kappa^\star,ij} = \left(  \frac{\partial^2  {\cal L}_{\rm post}(\kappa | {\cal D}) }{\partial \kappa_i \partial \kappa_j} \right)_{\kappa = \kappa^\star} \, .
\eea
For non-linear models as studied here these different steps need to be done numerically. Our use of the Laplace approximation was mainly motivated by the fact that it is much faster than MCMC techniques but still gave similar results in several tested cases. The Laplace approximation also appears more adapted to obtain a scalable algorithm that is applicable to thousands of load curves as done in this work (see Sec.~\ref{sec:numerical_tests}).

\subsection{Model selection}

In order to avoid over-fitting issues it is in general important to use a model selection procedure, e.g. to avoid using a modeling where cooling is used ($\alpha_c > 0$) for a building which display no cooling. In our setting this can be done by fitting various version of the modeling with one or several parameters set to $0$ (e.g. a model with no cooling and no impact of radiation, $\alpha_c = 0$ and $g = 0$, etc) and then comparing the {\it Bayesian information criterion} (BIC) of the different modelings \citep{rouchier2018solving}, defined as
\bea\label{eq:bic}
BIC = k \log(n_{train}) - 2 {\cal L}_{like} \, .
\eea
Other information criterion could be used (see e.g. \citep{mcelreath2018statistical}) but we do not try to compare them here.

\subsection{Long vs short terms predictions}

The introduction of autocorrelation in the modeling makes it possible to {\it improve short term predictions using past observations}. Indeed, using \eqref{eq:EQ_AR1_def}, the model \eqref{eq:ES_w_sun} can be rewritten as 
\bea
E_d = \rho E_{d-1} + f(X_d) - \rho f(X_{d-1})  + \eta_d \, ,
\eea
with $\eta_d$ a Gaussian white noise with std $\sigma'$. Iterating one gets $\forall m \geq 1$
\bea
E_{d+m} = \rho^m E_d + f(X_{d+m}) - \rho^m f(X_d) + \sum_{d' = d+1}^{d+m} \rho^{d+m-d'} \eta_{d'} \, .
\eea

Therefore the {\it $m$ steps ahead prediction of $E_{d+m}$ given $E_{d}$} is 
\bea
\hat{E}_{d+m}^{(m)} =   f(X_d) + \rho^m  E_{d} - \rho^m f(X_{d}) \, ,
\eea
with a variance $\sigma_m^2 = <(\hat{E}_{d+m}^{(m)}  -E_d)^2> = (\sigma')^2  \sum_{d' = d+1}^{d+m} \rho^{2(d+m-d')}  = (\sigma')^2  \frac{1 - \rho^{2m}}{1 - \rho^2}$. Since $| \rho | \leq 1$ the uncertainty (or variance of prediction $\sigma_m$) increases with $m$. As soon as $|\rho|^m \ll 1$ one can neglect the term involving $\rho$ in the above equations and one ends up with the long term prediction
\bea
\hat{E}_d^{(\infty)} =    f(X_d) \, ,
\eea
which has a variance $\sigma_\infty^2 = \frac{(\sigma')^2}{1-\rho^2}$. Note that when $\rho = 0$ (absence of autocorrelations) the distinction between long and short term predictions becomes immaterial. In the presence of autocorrelation the accuracy of the modeling  decreases with $m$. In the M\&V context considered in this work one is typically interested in the long-term prediction accuracy as savings are evaluated during a long time period.

\subsection{An example and addressing the instability issue on $\rho$}\label{subsec:instab}

We now focus on analyzing a specific example that uses EnergyPlus (E+) simulations of a modeling of a residential house that can be found on the web\footnote{We use the modeling of a residential house available from the prototype building models available on \citep{IECC_prototypes}.}. The goal of this example is not to assess the performance of the modeling in a general setting, which is the subject of the next sections, but rather to highlight an instability in the calibration of the parameter $\rho$ which was already known in the M\&V framework for linear models \citep{ruch1999prediction}.

\begin{figure}[h]
    \centering
    \includegraphics[width=0.45\textwidth]{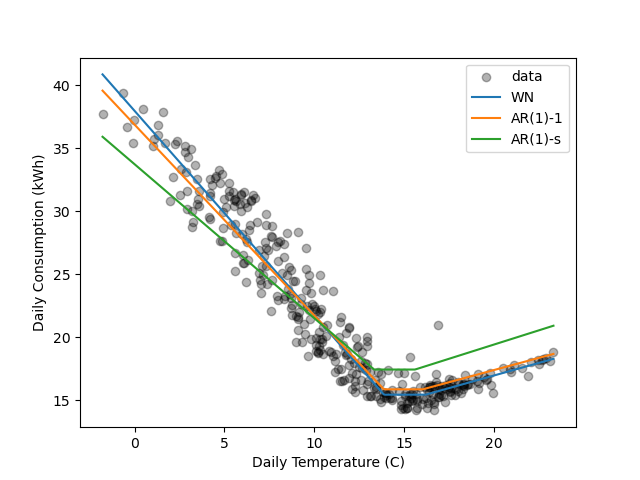}
    \caption{E+ simulation data together with the average predictions of the obtained energy signature modeling using three strategies (see text).}
    \label{fig:eplusES1}
\end{figure}

We show on Fig.~\ref{fig:eplusES1} the consumption vs temperature plot of the E+ modeling together with the fitted modeling using here three procedures 
\begin{itemize}
    \item {\bf White noise assumption (WN)}: here the parameter $\rho$ is constrained to $0$
    \item {\bf AR(1) with one-step calibration (AR(1)-1)}: here the modeling is first calibrated assuming $\rho = 0$. Then a value of $\rho = \rho_1$ is obtained by evaluating the autocorrelation at lag $1$ of the residuals of the modeling. The modeling is then recalibrated assuming $\rho = \rho_1$. This is akin to the Cochrane-Orcutt procedure interrupted after one step.
    \item {\bf AR(1) with simultaneous calibration (AR(1)-s)}: here $\rho$ is calibrated simultaneously with all parameters.
\end{itemize}
In all cases in Fig.~\ref{fig:eplusES1} we have also constrained $g = 0$ (i.e. a modeling using only the external temperature as a feature) for clarity of the exposure. On Fig.~\ref{fig:eplusES1} one can observe a phenomenon that, in our experience, often occur with M\&V data. When calibrating a modeling without constraining $\rho$, an instability can occur where $\rho$ increases close to $1$ and, although ${\cal L}_{post}$ increases, the predictive accuracy of the modeling apparently decreases. To understand what is happening, it is interesting to evaluate several metrics for the different modeling and to distinguish the long term and short term predictive performance as suggested in the previous section. In particular we consider the R-squared ($R2$) and coefficient of variation of the root mean squared error ($cv\_rmse$)  (see standard definitions in e.g. \citep{reddy2011applied}) for the long term predictions $\hat{E}_d^{(\infty)}$ and one step ahead predictions $\hat{E}_d^{(1)}$. The results are summarized in Table~\ref{table:1}.

 \begin{table}[h!]
\centering
\begin{tabular}{c| c | c | c  } 
 model & WN &  AR(1)-1 &   AR(1)-s \\ [0.5ex] 
 \hline\
  $\rho$ & 0  & 0.48 &  0.84  \\ 
  \hline
$R2^{(1)}$  & 0.94 & 0.95 & 0.95 \\
  \hline
$R2^{(\infty)}$  & 0.94 & 0.93 &  0.85 \\ 
  \hline
$cv\_rmse^{(1)}$  & 7.9 \% &  6.5 \% & 6.4 \%  \\ 
  \hline
$cv\_rmse^{(\infty)}$   & 7.9 \% & 7.9 \%  &  11.7 \%\\
  \hline
 ${\cal L}_{like}$   & -1526  & -1473 & -1463 \\ 
  \hline
${\cal L}_{post}$  & -1511 & -1458 & -1449 \\  [1ex] 
\end{tabular}
\caption{Fit metrics for the different Bayesian approaches considered in Sec.~\ref{subsec:instab}.}
\label{table:1}
\end{table}

\begin{figure}[h]
    \centering
    \includegraphics[width=0.45\textwidth]{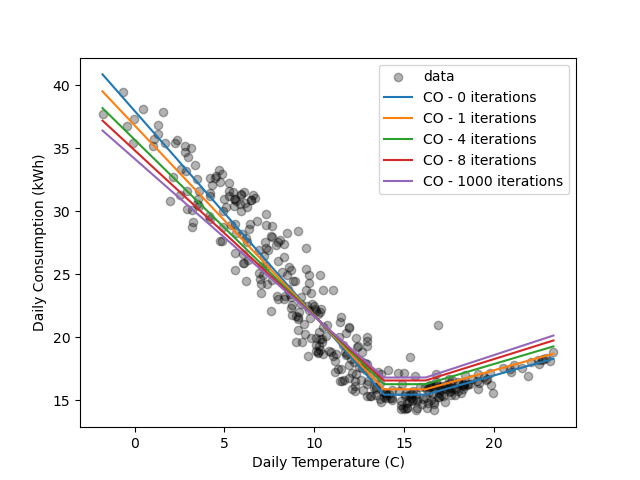}
    \caption{E+ simulation data together with the predictions of a linear model with hdd and cdd calibrated using the Cochrane-Orcutt (CO) procedure with a varying number of iterations $n_{C0}$.}
    \label{fig:eplusCO}
\end{figure}

One can observe that when calibrating $\rho$ simultaneously with other parameters (AR(1)-s) the obtained modeling has: (i) the largest log posterior and log-likelihood (and is therefore selected by the optimization procedure) ; (ii) the best performance metrics for the one-step predictions ; (iii) the worst metrics for the long time predictions. This seems to be a general issue with M\&V data which likely comes from the fact that temperature and consumption data are very autocorrelated with a value of $\rho$ close to 1. For practical application this is an unwanted behavior as in the M\&V context we are typically interested in the long term accuracy. This instability was already described in \citep{ruch1999prediction} in the context of linear models with autocorrelated errors when applying the Cochrane-Orcutt (CO) procedure. To confirm this we calibrate on the same data a linear regression model using heating and cooling degree days and the Cochrane-Orcutt procedure (see Sec.~\ref{subsubsec:ruch}) with a varying number of iterations  $n_{CO}$. The results are shown Fig.~\ref{fig:eplusCO} and in Table~\ref{table:2}. It is clear by comparing with the previous results that the same phenomenon is occurring: when the number of iterations increases, $\rho$ increases, the long term accuracy $R2^{(\infty)}$ decreases and the short term  accuracy $R2^{(1)}$ marginally increases ($R2^{(1)}(n_{C0} =1000) = 0.954$ and $R2^{(1)}(n_{C0} =1) = 0.952$).

\begin{table}[h!]
\centering
\begin{tabular}{c| c | c  | c} 
 $n_{CO}$ & $\rho$ & $R2^{(1)}$ & $R2^{(\infty)}$ \\ [0.5ex] 
 \hline\
 0 & 0 & 0.94  & 0.94  \\ 
  \hline
 1 & 0.47 & 0.95 & 0.93   \\
  \hline
 4 & 0.63 & 0.95  & 0.91  \\
  \hline
 8 & 0.70 & 0.95  & 0.90 \\
  \hline
 1000 & 0.75 & 0.95 & 0.89   \\ [1ex] 
\end{tabular}
\caption{Fit metrics for linear models calibrated using the Cochrane-Orcutt procedure with a varying number of iterations (see Sec.~\ref{subsec:instab}).}
\label{table:2}
\end{table}

In  \citep{ruch1999prediction}, as recalled in Sec.~\ref{subsubsec:ruch}, the solution to this instability which was proposed in the context of linear models was to use a hybrid modeling where model coefficients were estimated assuming $\rho =0$ but computing uncertainties by taking into account autocorrelations. This solution cannot strictly be adapted in our setting but in our case we simply propose to always use the "AR(1) with one-step calibration" as defined above\footnote{An alternative approach could be to use a strong prior on $\rho$ to avoid the instability. In this spirit the prior ${\cal L}_{\rm prior}(\rho) = -\frac{1}{(1-\rho)^2}$ was tested but still led to  less stable results compared with the one-step calibration.}. Although this treatment of the $\rho$ parameter should be viewed as an engineering trick rather than a rigorous Bayesian treatment, our research suggests that it  permits to take into account autocorrelation in a satisfying manner while maintaining an appropriate long term prediction accuracy (see Sec.~\ref{sec:numerical_tests}).

We note that the issue of the instability on $\rho$ was not addressed in the related work \citep{shonder2012Bayesian} treating of Bayesian uncertainties with AR(1) correlated errors in a M\&V context. On the theoretical side this is the main contribution of our paper to the issues of computing uncertainties for M\&V in a Bayesian setting.

\begin{figure}[h]
    \centering
    \includegraphics[width=0.45\textwidth]{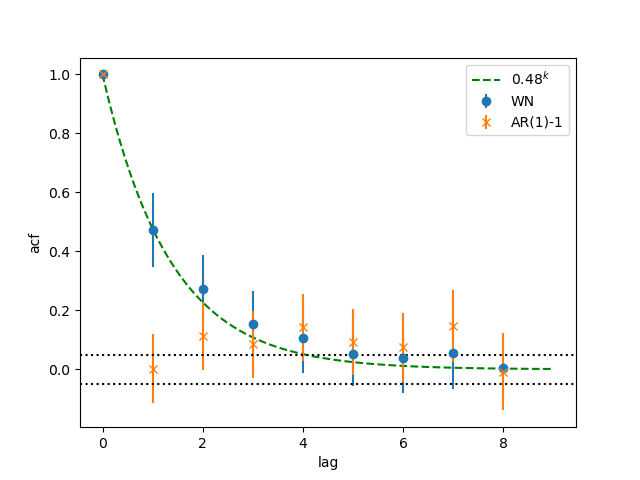}
    \caption{Autocorrelation function of the one-step prediction errors in the Bayesian energy signature model with white noise assumption (WN, blue dots) and AR(1) autocorrelated errors (AR(1)-1, orange dots). The green dashed line represents the autocorrelation function of a AR(1) process with $\rho = 0.48$. }
    \label{fig:acf}
\end{figure}

To conclude this section it is finally interesting to numerically compute the autocorrelation function of the $1$-step prediction errors for the WN and AR(1)-1 model to confirm that the assumed AR(1) structure represents the data well. The results are shown on Fig~\ref{fig:acf}. It can be seen that residuals in the WN model are autocorrelated with an ACF that is close to that of an AR(1) process. On the other hand the remaining autocorrelation of residuals of the 1 step predictions in the AR(1)-1 model are close to 0, confirming that the autocorrelation structure was satisfyingly modeled. Although the hypothesis is verified on this numerical example, let us emphasize that the AR(1) hypothesis probably breaks down in some cases. Investigating when this happens and how to deal with such cases is left for future work (see also \citep{rouchier5333168bayesian} for work in that direction).

\subsection{Computing uncertainties in M\&V  in a Bayesian setting}

Once a modeling has been selected using the $BIC$ criterion and the AR(1)-1 calibration procedure, uncertainties can be computed easily using the following. First, using the Laplace expression of the posterior \eqref{eq:laplace1} together with a numerical approximation of the Hessian \eqref{eq:hessian}, one can obtain samples of the posterior distribution of model parameters 
 $\kappa_i$, $i = 1, \cdots , n_{\rm samples}$ with $n_{\rm samples}$ a large number (we typically use $n_{\rm samples} = 10000$). Then for each $\kappa_i =  (bl_i , \alpha_{hi}, \theta_{hi}, g_i, \alpha_{ci}, \rho , \sigma'_i)$, a possible prediction during the observation period is obtained as $f(\kappa_i ; X_{obs}) + \epsilon_i$ where $\epsilon_i$ is a random realization of a stationary AR(1) noise with parameters $\rho, \sigma'_i$ with $n_{obs}$ observations. Then $f(\kappa_i ; X_{obs}) + \epsilon_i$ can be summed on the $n_{obs}$ observations to obtain a possible prediction for the total consumption during the observation period $\hat{E}_{tot,i}$. We therefore obtain a distribution of savings and any confidence interval\footnote{In Bayesian settings it is customary to speak of credible intervals.} can then be estimated by studying the distribution.

\subsection{Generalization to other non-linear models and influence of some features on the predicted total consumption}

\begin{figure}[h]
    \centering
    \includegraphics[width=0.45\textwidth]{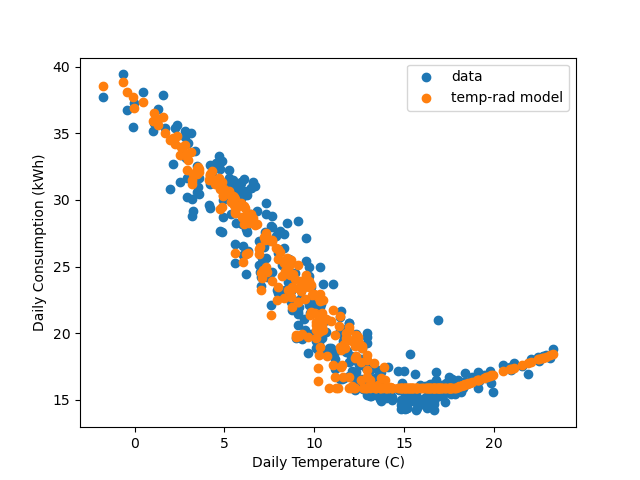}
    \caption{E+ simulation data together with the average predictions of the obtained energy signature modeling using external temperature and radiation as features.}
    \label{fig:eplustemprad}
\end{figure}

\begin{figure}[h]
    \centering
    \includegraphics[width=0.45\textwidth]{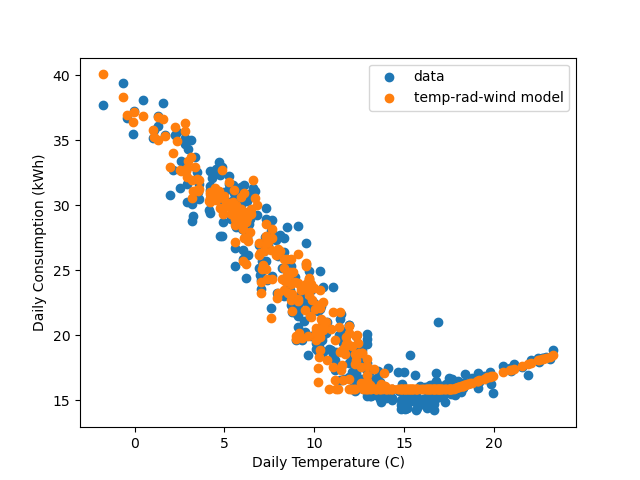}
    \caption{E+ simulation data together with the average predictions of the obtained energy signature modeling using external temperature, radiation and wind speed as features.}
    \label{fig:eplustempradwind}
\end{figure}

One of the main advantage of our approach is that it can be transparently adapted to any non-linear regression shape function $f(\kappa \, ; X_d)$. For example, following \citep{rasmussen2020method}, we can introduce in \eqref{eq:ES_w_sun} a coupling of heat loss in the winter season with the wind-speed $v_w$ as

\bea
 \hspace{-0.5cm} f(\kappa \, ; X_d) & =&  bl +
 \alpha_h \times (1+ \alpha_w \times v_w) \times (\theta_h - \theta_d - g\times I_{d})^{(+)}   \, \nonumber \\
 &+& \alpha_c \times (\theta_d - \theta_c)^{(+)} \, ,
 \label{eq:ES_w_sun_and_wind} 
\eea
where a new parameter $\alpha_w$ was introduced. For comparison with the model with temperature only, we show on Fig.~\ref{fig:eplustemprad} and Fig.~\ref{fig:eplustempradwind} the results of calibrating a modeling including radiation (model \eqref{eq:ES_w_sun}) and a modeling including radiation and wind-speed (model \eqref{eq:ES_w_sun_and_wind}), still using the same synthetic data. It can be seen that the fit increases when adding features in the modeling (and indeed features are estimated as being relevant according to the BIC criterion of \eqref{eq:bic}).

\begin{figure}[h]
    \centering
    \includegraphics[width=0.45\textwidth]{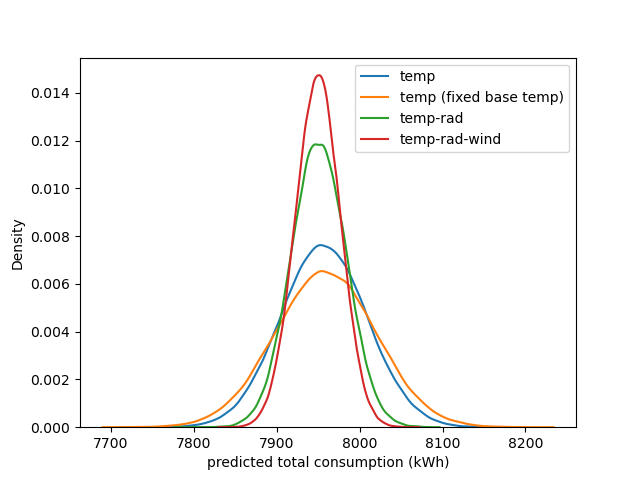}
    \caption{Comparison of the predicted distributions of annual consumption in the Bayesian energy signature model with AR(1) errors (AR(1)-1) in several cases: using only the external temperature as a feature (blue line) ; using only the external temperature as a feature with fixed base temperature (orange line)  ; using the external temperature and radiation as features ; using the external temperature, radiation and wind speed as features.}
    \label{fig:eplusFSU}
\end{figure}

In our setting, the most interesting aspect is to assess the effect of adding features on the prediction of the total consumption during some observation period. This is done in Fig.~\ref{fig:eplusFSU} where we show the predicted $\hat{E}_{tot}$ (using here for simplicity the same climate data as for the training period) for the various combination of features (temperature only, temperature and radiation, temperature and radiation and wind speed). In addition we also display on Fig.~\ref{fig:eplusFSU} the results obtained when neglecting the uncertainties on the base temperatures (for a modeling with only temperature as a feature), which is the usual assumption made in the M\&V field. On this simple example we see that adding features decreases the uncertainty, as more and more noise gets in fact explained by the modeling. On this example neglecting the uncertainties on the base temperature seem to only have a weak effect, confirming the usual assumption. Obviously the results shown here are only valid for this example and should not be considered as general conclusions. The goal of the next sections is to test our approach more thoroughly on a large scale dataset and to compare it with other approaches to uncertainty quantification reviewed in Sec.~\ref{sec:overview}.

\section{Constructing a dataset adapted to test M\&V workflows} \label{sec:jdds}

In order to test M\&V approaches, we construct a synthetic dataset containing the results of stochastic dynamic thermal simulation of thousands of buildings.

To obtain a representative set of building characteristics we use the French DPE-AUDIT Observatory (\url{https://observatoire-dpe-audit.ademe.fr}). It is a national platform managed by ADEME (the French Agency for Ecological Transition) that provides open-access data related to Energy Performance Certificates (called DPE in France) and Energy Audits for residential and tertiary buildings. From the most recent DPEs (following the July 2021 reform), a dataset of 3000 DPEs at the residential unit scale was extracted.

The building characteristics of interest contained in the DPE include the living surface area, geographical coordinates, total surface area, orientation, and thermal properties of all walls, windows/doors as well as the roof and the floor. The types of energy source and the systems used for heating, domestic hot water, cooling and ventilation are also available. Furthermore, the size of the photovoltaic system is included, if present.

These characteristics are then used to construct for each building a single-zone thermal model in the energy simulation platform DIMOSIM (DIstrict MOdeller and SIMulator). This platform was developed by CSTB (the Scientific and Technical Center for Building) and integrates a stochastic occupancy profiles generator based on a 2010 survey conducted by INSEE (the French National Institute of Statistics and Economic Studies). The generator determines the number of occupants and their status (e.g., student, worker, retirees…) according to the living surface area and derives for each of them a stochastic 10-min step activity profile among 24 activities. Each activity (e.g., at work, cooking, hygiene, sleeping, holidays) is associated with specific equipment loads and corresponding water draw-off profiles. 

Additionally, blinds are controlled during the day in response to the operative temperature within the thermal zone. For each building, a random start and end day, along with comfort, economic and holiday temperature setpoints for the heating and cooling seasons, are selected prior to running simulation. Window opening scenarios and lighting scenarios are not included in the current simulation framework. See more information on DIMOSIM modeling tool and the generation of stochastic occupancy profiles in \citep{garreau2021district}.

\begin{figure}[h]
    \centering
    \includegraphics[width=0.35\textwidth]{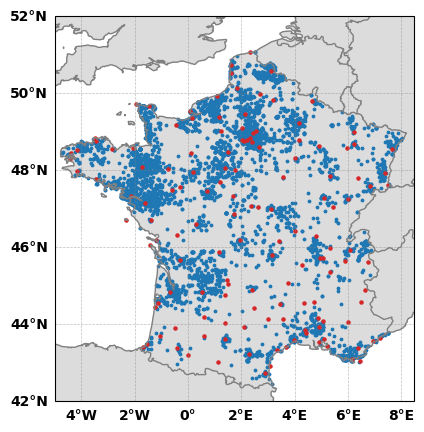}
    \caption{Coordinates of simulated buildings (blue dots) and weather stations (red dots).}
    \label{fig:frenchmap}
\end{figure}

Because the building archetypes (e.g., envelope properties, facade orientation) are consistent with local climate conditions, the geographically closest weather file is selected for each building from a set of 152 Météo-France (the French national meteorological and climatological service) weather stations. Weather files have been created in .epw format using historical in-situ hourly data for temperature, humidity, atmospheric pressure, and wind from the weather stations and are completed with solar radiation data at the station coordinates obtained from \citep{CAMS2020}. For each station location, a six-years hourly weather file (covering 2017 to 2022) has been created. 

We show on Fig.~\ref{fig:frenchmap} a map illustrating the geographical dispersion of the simulated buildings across France, along with the Météo-France weather station network used in the weather files.

\begin{figure}[h]
    \centering
    \includegraphics[width=0.45\textwidth]{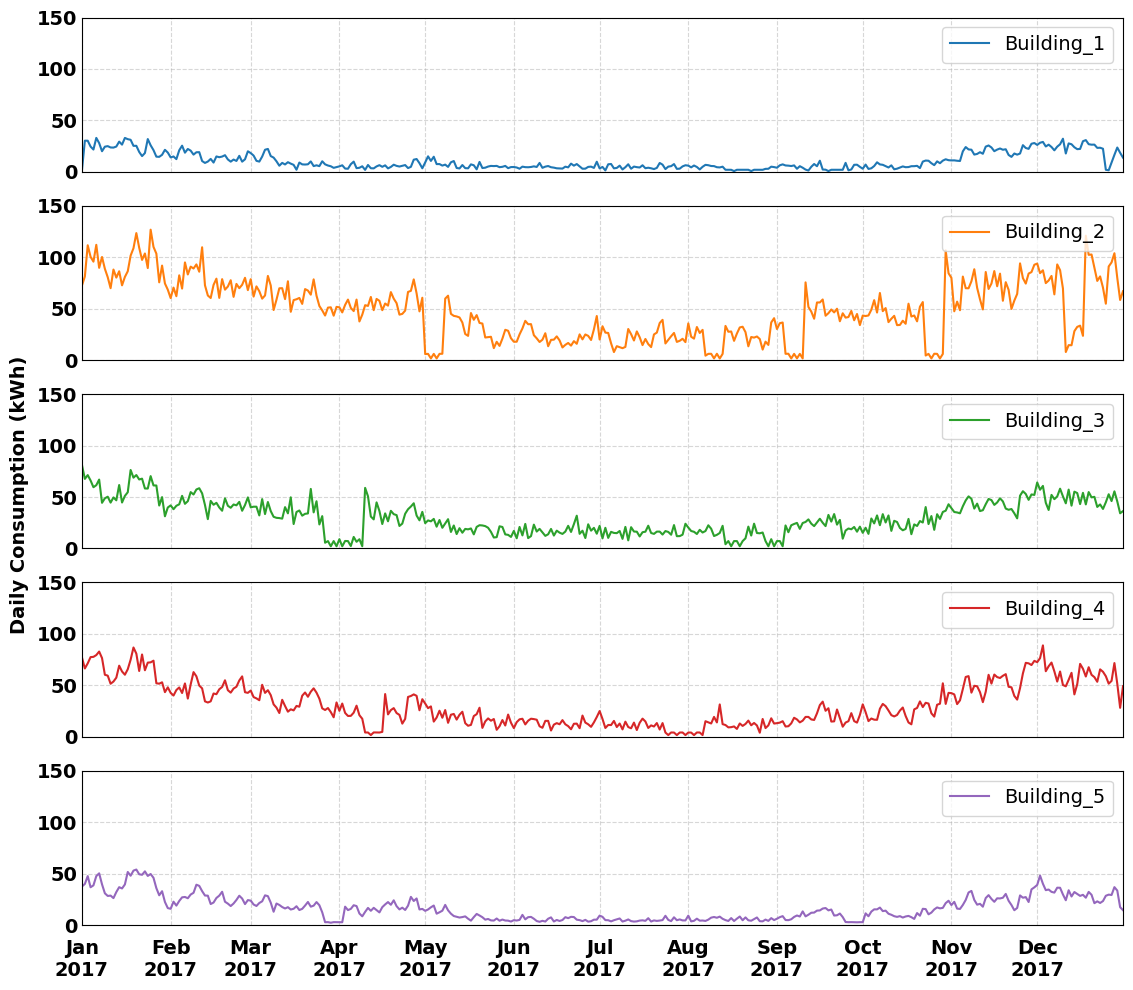}
    \caption{Total (sum of all energy sources) daily energy consumption resulting from DIMOSIM simulations for 5 buildings of the dataset over one year.}
    \label{fig:datasetsubplot}
\end{figure}

\begin{figure}[h]
    \centering
    \includegraphics[width=0.45\textwidth]{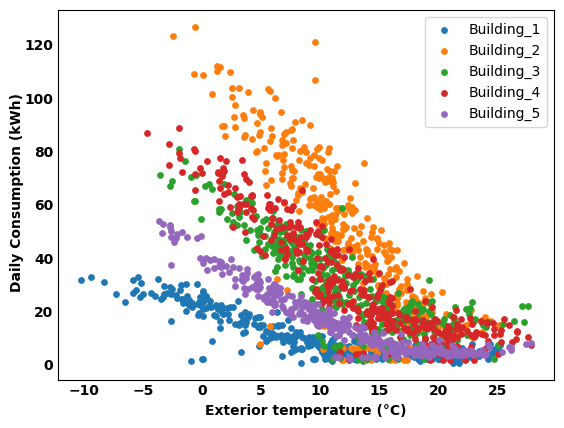}
    \caption{Energy signature for the 5 buildings of Fig.~\ref{fig:datasetsubplot}.}
    \label{fig:es_5buildings}
\end{figure}

The resulting dataset provides hourly energy consumption data for various energy sources (electricity, gas, biomass, oil and photovoltaic) over the six-years simulation period covering all 3000 buildings. The Fig.~\ref{fig:datasetsubplot} shows a subset of five randomly selected buildings from the dataset for the year 2017. One can observe the disparity in simulated consumption between these buildings. In particular, the holiday periods differ from one building to another, thanks to the stochastic profile generator used to simulate occupant activities. For these buildings, the energy signature profiles are presented in Fig.~\ref{fig:es_5buildings} and show the disparity between buildings behaviors.


\section{Test of M\&V approaches}\label{sec:numerical_tests}

\subsection{Test protocol}

In order to test M\&V algorithms (including modeling and uncertainty computation approach) we perform for each building in the dataset and each algorithm the following tasks:
\begin{enumerate}
    \item For each complete year of data $y_{pre} \in [2017, 2022]$, calibrate the modeling (6 possibilities).
    \item Predict with the modeling the theoretical consumption + uncertainties interval for all other years $y_{post} \in [2017, 2022], y_{post} \neq y_{pre}$ (5 possibilities)
    \item Compare the theoretical consumption predicted by the statistical approach with the real one.
\end{enumerate}
Hence, performing this workflow for each building in the dataset, we get $6 \times 5 \times N_b$ experiments per algorithm (here $N_b =3000 $ denotes the number of buildings). We thus simulate $90000$ M\&V experiment for each algorithm. This permits to obtain for each tested algorithm accuracy scores that we will explain. To simplify our test we also always apply the above strategy on the total consumption of the building (ex: electricity + gas for a gas-heated building) instead of doing it fluid by fluid. Below we also present results where vacation days are filtered out of the dataset for clarity. Similar results are obtained by using standard outlier filtering techniques based e.g. on the analysis of a z-score.

\subsection{Summary of tested modelings and approaches (daily data)}

Let us summarize clearly the modeling and uncertainty computation approaches that are tested here.

\medskip

{\bf Mean day model } In order to have a simple baseline model that can be used to put in perspective the results of more sophisticated approaches, it is useful to define a simple naïve model and to study its performances. We use here the mean day model as a baseline: predictions are simply made by using the average daily consumption of the training period: $\hat{E}_t  = < E_{t',train}>$ and uncertainty are computed assuming that residuals are Gaussian and uncorrelated. This is therefore a modeling where no climate correction is applied.

{\bf Linear regression with constant base temperatures} The simplest approach that is considered is a linear regression model taking as features heating and cooling degree days computed with base temperatures of $18 °C$ and $22°C$ as in \eqref{eq:5points}. For this modeling we test 5 approaches to uncertainties computation:
\begin{enumerate}
    \item Standard OLS: exact matrix calculations neglecting autocorrelations as in Sec.~\ref{subsubsec:ols}.
    \item Ruch: matrix calculations using the results of \citep{ruch1999prediction} as presented in Sec.~\ref{subsubsec:ruch}.
    \item Newey-West: matrix calculations using Newey-West estimators as presented in Sec.~\ref{subsubsec:newey-west}.
    \item ASHRAE 14: approximate formula presented in Sec.~\ref{subsubsec:ashrae14}.
    \item Conformal predictions: conformal predictions approach using cross validation as presented in Sec.~\ref{subsec:cv}.
\end{enumerate}
The 5 approaches will be respectively denoted as $LR(18,22) - OLS$, $LR(18,22) - Ruch$, $LR(18,22) - NW$, $LR(18,22) - ASHRAE 14$, $LR(18,22) - CV$.

\medskip

{\bf Energy Signature} The second approach is the energy signature with or without taking into account radiation as in \eqref{eq:ES1} (denoted $ES$) or \eqref{eq:ES2} (denoted $ES - sun$). For these modelings we apply the Bayesian approaches using the white noise approximation ($Bayes - WN$) or AR(1)-1 approximation ($Bayes - AR(1)$). For comparison we also test the ASHRAE 14 formula \eqref{eq:ashrae14}. We therefore end up with $6$ approaches $ES - Bayes - WN$, $ES - Bayes - AR(1)$, $ES -  ASHRAE 14$, $ES - sun - Bayes - WN$, $ES - sun - Bayes - AR(1)$, $ES - sun- Bayes - ASHRAE 14$.

\medskip

{\bf Other tested approaches} It's also interesting to extend our test to more complex machine learning approaches and to use the ASHRAE 14 formula \eqref{eq:ashrae14}. Indeed, using this formula for any model is today the standard practice of the industry. We consider two different modelings. The first is the Caltrack approach that seems to be more and more widely used in the US. We use the open-source Python implementation available using the OpenEEmeter package (version 4.1). The Caltrack approach is very close to the energy signature except that several energy signatures can be used to model a single building, depending on the season. It uses only the external temperature as a feature. The second approach is {\it multi adaptative regression spline} (MARS) which is a machine learning regression technique. This algorithm appeared particularly well suited to M\&V in a numerical test bench of M\&V workflow including 11 algorithms \citep{osso2024}. As in \citep{osso2024} we use the open source python implementation available in the py-earth package. It is very versatile in the sense that it can be calibrated to represent any arbitrary complex continuous function. In this work we will consider applying MARS using either only the external temperature as a feature (MARS), or both external temperature and radiation (MARS - Sun). For our test of uncertainty quantification approaches both Caltrack and MARS have the interesting feature of in the end calibrating an analytic non-linear function. In both cases it is therefore possible to define a number of calibrated parameters to be used as input of the formula \eqref{eq:ashrae14}. For MARS we also test the use of conformal predictions. In Summary we test 5 other approaches: $Caltrack - Ashsrae14$, $MARS - ASHRAE 14$, $MARS - Sun- ASHRAE 14$, $MARS - CV$, $MARS - Sun- CV$.

\subsection{Summary of metrics considered }

For each algorithm, building and year of tested data we get to compare the true building consumption $E_{{\rm tot}}$ with the pointwise prediction of the algorithm $\hat{E}_{{\rm tot}}$ and the lower and upper bounds of a $95\%$ confidence interval $\hat{E}_{{\rm low}}$ and $\hat{E}_{{\rm top}}$. To test the accuracy of the prediction given by the confidence interval we consider three metrics: the internal coverage probability (ICP), the interval width (IW) and the Winkler score.

\begin{itemize}
    \item  
    The internal coverage probability measures the fraction of cases such that the true building yearly consumption lies in the 95\% predicted confidence interval. I.e we count a score of $1$ when $E_{{\rm tot}} \in [\hat{E}_{{\rm low}}, \hat{E}_{{\rm top}}]$, and $0$ otherwise. The ICP is obtained by averaging over buildings. Here it is expected that a perfect method should have $ICP \simeq 95\%$ by definition.
    \item 
    The interval width (IW) measures the width of the prediction interval, expressed in \% of the true consumption: $IW = \frac{\hat{E}_{{\rm top}}-\hat{E}_{{\rm low}}}{\hat{E}_{{\rm tot}}}$. Here one expects an accurate method to have the smallest possible $IW$.
    \item 
    Finally, the Winkler score tries to unite both ICP and IW in a single metric. It is defined as $Winkler = IW$ for cases such that the true consumption lies in the prediction interval, and $Winkler = IW + \frac{2}{\alpha} \Delta $ with $\Delta$ the distance (in \%) between the true consumption and the edge of the prediction interval otherwise. The best algorithm according to this metric score should have the smallest possible Winkler score.
\end{itemize}

In addition to these metrics we also keep track of other metrics tracking the pointwise accuracy and the goodness of fit of the algorithms:
\begin{itemize}
    \item The average error in $\%$, $me = <\frac{\hat{E}_{{\rm tot}} - E_{{\rm tot}}}{E_{{\rm tot}}} > $.
    \item The mean absolute error in $\%$, noted $mae = <\frac{|\hat{E}_{{\rm tot}} - E_{{\rm tot}}|}{E_{{\rm tot}}} > $.
    \item The value of the ACF at lag $1$ noted $\rho$ as in the core of the paper.
    \item The $R2$ and $CV(RMSE) $ of the modeling during the train and test period, $R2_{train}$, $R2_{test}$, $CV(RMSE)_{train}$ and  $CV(RMSE)_{test}$.
\end{itemize}

Finally we also show for informative purposes the average duration $ \Delta t $ taken by the algorithm\footnote{The duration includes model calibration, predictions and calculation of uncertainties, as measured on the laptop used to perform the simulation.}.

\subsection{Results and discussion}

The results obtained for all algorithms are summarized in Table~\ref{tab:rotated_metrics}.

\begin{sidewaystable}[htbp]
    \centering
\begin{tabular}{| c| c | c  | c | c | c | c | c  | c  | c  | c  | c | } 
\hline
 {\bf Algorithm} &  $ICP \, (\%)$ &  $IW \, (\%)$ & $Winkler \, (\%)$ & $me \, (\%) $  & $mae \, (\%) $ & $\rho$  &  $R2_{train}$ &  $R2_{test}$ & $CV(RMSE)_{train} \, (\%)$ & $CV(RMSE)_{test} \, (\%)$ & $ \Delta t \, (s) $  \\ [0.5ex] 
 \hline\
 $Mean Day$ & 63 & 13.6  & 62  & -0.25  & 5.86  & 0.84  & 0    & -0.01  & 63  & 64 & 0.00 \\ 
 \hline
 \hline \
 $LR(18-22) - OLS$ & 89 & 7.54   & 12.76  & -0.05  & 1.86  & 0.31  & 0.81    & 0.81  & 24  & 25 & 0.01 \\ 
 \hline\
 $LR(18-22) - Ruch$ & 97 & 11.0   & 12.38  & .  & .  & .  & .  & .  & .  & . & 0.08  \\ 
 \hline\
 $LR(18-22) - NW$ & 96 & 10.16   & 11.71  & .  & .  & .  & .   & .  & .  & .& 0.02 \\ 
 \hline\
 $LR(18-22) - ASHRAE 14$ & 97 & 15.14   & 20.24   & .  & .  & .  & .  & .  & .  & . & 0.03 \\
 \hline\
 $LR(18-22) - CV $ & 88 & 7.57   & 12.91  & .  & .  & .  & .   & .  & .  &  . & 0.16 \\ 
 \hline
 \hline\
 $ES - Bayes - WN$ & 92 & 6.99   & 14.08  & -0.01  & 1.55  & 0.25  & 0.83   & 0.83  & 23  & 23 & 0.19\\ 
 \hline\
 $ES - Bayes - AR(1)$ & 97 & 9.21   & 10.66   & .  & .  & .   & .  & .  & . & . & 0.20 \\ 
 \hline\
 $ES - ASHRAE 14$ & 97 & 12.92   & 19.95   & .  & .  & .  & .  & .  & .  & . & 0.05 \\
 \hline
 \hline\
 $ES - Sun -  Bayes - WN$ & 93 &	5.96 & 8.89 & -0.11	& 1.26 & 0.08 & 0.87 & 0.86	 & 20	& 20 & 0.20 \\ 
 \hline\
 $ES - Sun -  Bayes - AR(1)$ & 94 & 6.62   & 8.48  & .  & .  & .  & .   & .  & .  & . & 0.22 \\ 
 \hline\
 $ES - Sun -  ASHRAE 14$ & 93 & 6.91   & 8.89   & .  & .  & .  & .   & .  & .  &  . & 0.08\\
 \hline
 \hline\
 $Caltrack -  ASHRAE 14$ & 92	& 9.81 & 13.65 & -0.29	& 1.99	& 0.19	& 0.85	& 0.84 & 	21	& 22  & 7.34\\
 \hline
 \hline\
 $MARS -  ASHRAE 14$ & 96	 & 11.25 & 12.59 & 0.03 & 1.62 & 0.24 &  0.84 & 0.83 & 23	& 23 & 0.08 \\
 \hline\
 $MARS -  CV$ & 88 & 6.67 & 11.13 & .  & .  & .  & .   & .  & .  & . & 0.35 \\
 \hline
 \hline\
 $MARS - Sun - ASHRAE 14$ & 91 & 5.79	 & 8.47	& 0.00 & 1.20 & 0.04 & 0.88 & 0.87 & 18	& 19 & 0.10\\
 \hline\
 $MARS - Sun - CV$ & 85 & 4.37 & 9.14  & .  & .  & .  & .   & .  & .  & . & 0.51\\
 \hline
\end{tabular}
    \caption{Performance metrics for all tested algorithms. A dot means a value equal to the line above. Algorithms where grouped according to the shape of the function $f(X_t)$.}
    \label{tab:rotated_metrics}
\end{sidewaystable}

\begin{figure}[h]
    \centering
    \includegraphics[width=0.45\textwidth]{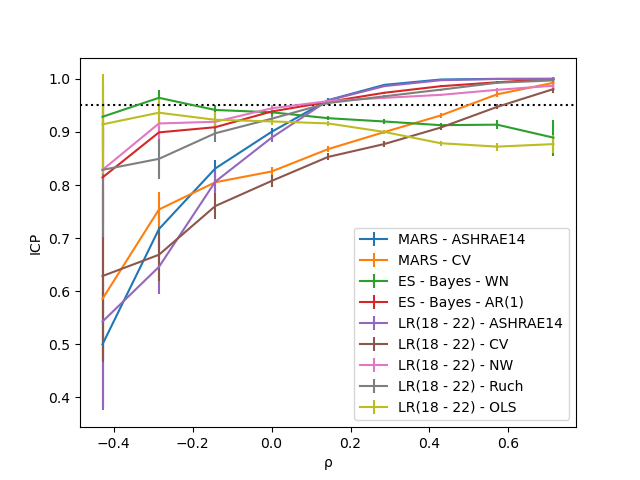}
    \caption{Measured evolution of the internal coverage probability as a function of the autocorrelation parameter $\rho$.}
    \label{fig:ICP}
\end{figure}

\begin{figure}[h]
    \centering
    \includegraphics[width=0.45\textwidth]{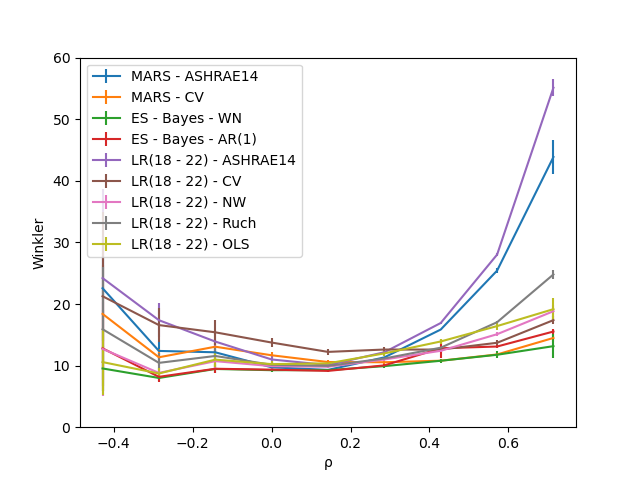}
    \caption{Measured evolution of the   Winkler score as a function of the autocorrelation parameter $\rho$.}
    \label{fig:Winkler}
\end{figure}

For a more complete interpretation of the results it is also interesting to look at the evolution of some of these metrics with the autocorrelation parameter $\rho$. This is done in Fig.~\ref{fig:ICP} and Fig.~\ref{fig:Winkler} for the ICP and the Winkler score (only for a subset of algorithm for readability).

The following observations can be made on the results of Table~\ref{tab:rotated_metrics}:
\begin{enumerate}
    \item Neglecting autocorrelations (LR(18-22) - OLS and ES - Bayes - WN) generally leads to too small ICP and the performance worsens as $\rho$ deviates from $0$ (see Fig.~\ref{fig:ICP} and Fig.~\ref{fig:Winkler}).
    \item While the ASHRAE 14 approach can lead to good overall metrics it can be seen on Fig.~\ref{fig:ICP} and Fig.~\ref{fig:Winkler} that its performance also greatly varies with the value of the autocorrelation parameter $\rho$. The conformal prediction approach display a similar behavior, although its performance seems lower than the ASHRAE 14 approach, with a consistent underestimation of the width of the confidence interval.
    \item The most robust techniques (on average and for various $\rho$, see Fig.~\ref{fig:ICP} and Fig.~\ref{fig:Winkler}) appears to be the Bayesian approach (AR(1)) for non-linear models and the Newey-West approaches for linear models.  These methods provides confidence intervals with an ICP consistently close to $95\%$ and a stable Winkler score. This should be compared with the results of \citep{touzani2019evaluation} with real world data where the best algorithm has an ICP of $71\%$ only.
    \item The most accurate algorithms appear to be MARS - Sun and ES - Sun, as measured by the $mae$ ($1.2\%$ and $1.26\%$). This should be compared to the value of $5.86\%$ obtained without climate correction (Mean Day model). While MARS - Sun appears as the most accurate, the uncertainty approaches that can be applied for this model do not lead to fully satisfying results.
    \item Including sun radiation improves the performance of the tested algorithms (MARS and energy signature) but the improvement remains modest. Still, one can see that the performance improvement obtained from using the Bayesian energy signature with radiation approach instead of using a linear regression with fixed base temperatures and using the ASHRAE 14 formula is quite important. Indeed the width of the prediction interval is almost three times smaller with the energy signature approach but the internal coverage probability remains close to 95\%. This is clearly useful to accurately measure energy savings, especially when energy savings effects are expected to be small.
    \item The main effect associated with neglecting the uncertainty in the base temperatures (as for linear models approaches) is to decrease the accuracy of the modeling ($mae$ and $IW$ increases compared with energy signature approaches), but the computed uncertainties still seem consistent with numerical results ($ICP$ close to $95\%$).
\end{enumerate}

Hence the study of these metrics provide answers to the main questions raised in the beginning of this work (see Sec.~\ref{subsec:summary_of_questions}). Let us now draw a general conclusion on our findings and discuss possible new research directions.

\section{Conclusion}\label{sec:conclusion}

We have reviewed in this work different techniques available for the computation of uncertainties in a M\&V context: analytic techniques available for linear models ; Bayesian techniques that can be applied to linear and non-linear regression models ; the ASHRAE 14 formula ; conformal prediction approach based on resampling. Among these techniques, the last two have the advantage of being applicable to any modeling, but their accuracy is uncontrolled and was tested here in a few cases. We have also improved in this work how autocorrelations can be taken into account in the Bayesian approach in a way that is adapted to M\&V data. Finally the different approaches were numerically evaluated on a large scale synthetic dataset containing 6 years of stochastic simulation data for 3000 buidings randomly chosen in France.

Compared with \citep{touzani2019evaluation}, which performed a similar test using real data with results indicating that current techniques fail at providing a consistent estimates of uncertainties, we arrive to a more nuanced picture\footnote{In the case of data with daily time resolution, which was the scope of this work.}. Indeed we find that
\begin{itemize}
    \item For linear models, analytic approaches such as \citep{ruch1999prediction} and the Newey-West \citep{newey1986simple} approach leads to good results.
    \item For non-linear regression model the Bayesian approach developed in this work leads to good results.
    \item On average the approximate formula of ASHRAE 14 \citep{ashrae14} provides satisfying results, although its accuracy quickly decreases when autocorrelations increase. A similar pattern is observed for the approach using cross validation but with overall poorer performances.
\end{itemize}

Overall our results show that for linear and non-linear regression models, linear and Bayesian approaches provide a satisfying way to estimate the uncertainties using well grounded theoretical principles. Other approaches such as the ASHRAE 14 formula provide a decent approximation, but its performance can quickly decrease e.g. in the presence of high autocorrelations. For industry practitioners, it might be the case that the ASHRAE 14 formula provides a "good enough" approximation to the true uncertainty interval. Its simplicity and clarity remains an important benefit. However, if more rigorous statistical bounds are needed, other approaches such as ours can lead to more accurate results.

Besides these results our work also shows in a controlled context the gains in accuracy that can be expected from several modeling improvements: (i) gains obtained from using fixed base temperatures to using base temperatures that are optimized at the building level as in the energy signature approach ; (ii) gains obtained from using a modeling that uses only the external temperature as a feature to a modeling using other climate features. This should help in a practical setting to decide whether or not the use of sophisticated methods is justified.

While our work suggests that for linear and non-linear regression model with daily time resolution some existing techniques permit to estimate the uncertainty accurately, there are many directions in which this study could be extended. 
\begin{enumerate}
    \item 
First, staying in the context of energy data with a daily time resolution, it would be interesting to study further the applicability of these techniques using real world data. In particular it would be interesting to check if the simplifying hypothesis of AR(1) autocorrelations is sufficient in practice (see also \citep{rouchier5333168bayesian} for a study in that direction). Studying in details the dataset used in \citep{touzani2019evaluation} would be interesting to try to understand if the negative results obtained there truly show some limitations of the tested techniques, or if the results simply show that the dataset cannot be used to estimate reliably the accuracy of uncertainty quantification approaches.  
\item
Second, it would be interesting to study how the techniques behave when using data with hourly time resolution, e.g. focusing first on the time of week and temperature model of \citep{mathieu2011quantifying}.
\item
Finally it would be interesting to find techniques applicable to modelings that cannot be written as non-linear regression models, e.g. to compute uncertainties when using approaches based on decision trees as in \citep{touzani2018gradient}.
\end{enumerate}

\section*{Acknowledgement}
We thank our colleagues at CSTB for numerous discussions and in particular Etta Grover-Silva, Enora Garreau and Jean-Baptiste De Fouchier for help in setting up the DIMOSIM simulations, and Ugo De Filippis for collaborations on related topics. This research was funded by the Watt Watchers program, funded in France by the "dispositif des certificats d’économies d’énergie".




\bibliographystyle{elsarticle-harv} 
\bibliography{refs}






\end{document}